\documentclass[letterpaper,journal]{IEEEtran}
\usepackage{amsmath,amsfonts}
\usepackage{algorithmic}
\usepackage{algorithm}
\usepackage{array}
\usepackage[caption=false,font=normalsize,labelfont=sf,textfont=sf]{subfig}
\usepackage{textcomp}
\usepackage{stfloats}
\usepackage{url}
\usepackage{verbatim}
\usepackage{graphicx}
\usepackage{subcaption}
\usepackage{cite}
\usepackage{amssymb}
\usepackage{booktabs}
\usepackage{multirow}
\usepackage{xcolor}
\begin{document}

\title{A Markov Decision Process Framework for Enhancing Power System Resilience during Wildfires under Decision-Dependent Uncertainty}

\author{Xinyi Zhao,~\IEEEmembership{Student~Member,~IEEE,} Prasanna Raut,~\IEEEmembership{Student~Member,~IEEE,} Chaoyue Zhao, \IEEEmembership{Member,~IEEE,} Alexandre Moreira, \IEEEmembership{Senior Member,~IEEE} }

\markboth{Journal of \LaTeX\ Class Files,~Vol.~14, No.~8, August~2021}%
{Shell \MakeLowercase{\textit{et al.}}: A Sample Article Using IEEEtran.cls for IEEE Journals}


\maketitle

\begin{abstract}
Wildfires pose an increasing threat to the safety and reliability of power systems, particularly in distribution networks located in fire-prone regions. To mitigate ignition risk from electrical infrastructure, utilities often employ safety power shutoffs, which proactively de-energize high-risk lines during hazardous weather and restore them once conditions improve. While this strategy can result in temporary load loss, it helps prevent equipment damage and wildfire ignition development in the system. In this paper, we develop a state-based decision-making framework to optimize such switching actions over time, with the goal of minimizing total operational costs throughout a wildfire event. The model represents network topologies as Markov states, with transitions influenced by both exogenous weather conditions and endogenous power flow dynamics. To address the computational challenges posed by the large state and action spaces, we propose an approximate dynamic programming algorithm based on post-decision states. The effectiveness and scalability of the proposed approach are demonstrated through case studies on 54-bus and 138-bus distribution systems, showcasing its potential for enhancing wildfire resilience across different grid configurations.
\end{abstract}


\vspace{-0.4cm}
\section{Introduction}
\vspace{-0.1cm}
Wildfire risk is rising worldwide as climate patterns shift \cite{bowman2017human}, creating increasing threats to communities and critical infrastructure. Recent events, such as the Southern California wildfires of January 2025, have caused tens of billions of dollars in damage and prompted widespread preventive power shutoffs \cite{seydi2025assessment}. Power systems are particularly exposed: wildfires can destroy key grid assets, while electrical faults have ignited some of the most destructive fires on record. In California, for instance, nearly half of the 20 most catastrophic wildfires in recent decades were traced to power line failures \cite{vazquez2022wildfire}, underscoring both the severity of the problem and the dangerous feedback loop between an aging grid and escalating wildfire hazards.

To reduce the likelihood of power line–caused ignitions, utilities in high-risk regions have increasingly implemented Public Safety Power Shutoff (PSPS) programs \cite{huang2023review}. Under PSPS, selected lines are proactively de-energized during periods of extreme weather conditions characterized by high winds, low humidity, and abundant dry vegetation to eliminate electrical ignition sources. Although effective in reducing fire risk, this strategy imposes significant reliability costs, as outages may persist for many hours or days \cite{abatzoglou2020population}. PSPS therefore presents a challenging operational trade-off: accepting controlled outages to prevent potentially catastrophic wildfire events.


Determining how and when to implement PSPS poses complex sequential decision-making challenges \cite{lesage2023optimally}. Operators must decide which lines to shut off and when to restore them as conditions evolve, balancing reductions in ignition risk against the societal and operational impacts of outages. These decisions must account for uncertainties in weather, wildfire progression, and system loading, which depend both on exogenous factors (e.g., meteorological forecasts) and endogenous factors (e.g., network topology changes following switching actions) \cite{moreira2024distribution}.

In response to these challenges, an emerging body of research has sought to support power system operations under wildfire threat. Existing work \cite{abdelmalak2022enhancing, moreira2024distribution, pianco2025decision} includes preventive dispatch and switching strategies for transmission resilience formulated via Markov decision processes, integrated emergency management models combining generator redispatch and load shedding, and distribution-level tools for coordinating microgrids or mobile generators to sustain service during wildfires. Additional efforts \cite{wang2019markov, wang2020mdp, abdelmalak2021markov} examine network reconfiguration and islanding strategies to harden distribution systems against natural hazards. While these studies advance operational resilience, most focus on maintaining system functionality or protecting assets rather than on the explicit timing and scope of preemptive de-energization. As a result, the sequential decision problem of when and where to shut off or restore lines in response to evolving wildfire risk remains insufficiently addressed in the literature.

This paper addresses this gap by developing a state-based dynamic decision framework for optimal PSPS scheduling in distribution networks. The network’s operational configuration is represented as a state in a Markov Decision Process (MDP), defined by the current topology and associated system status, while actions correspond to switching decisions that energize or de-energize selected lines. State transitions capture the combined influence of exogenous wildfire-related conditions and endogenous power-flow changes. The objective is to minimize the total operational cost over a wildfire event, which includes the cost of active power purchased at substation nodes, the cost of unserved energy, and costs associated with switching actions. Because the penalty for unserved load is significantly higher than the cost of energy purchases, the model prioritizes maintaining service wherever it is safe to do so.

Solving the resulting MDP exactly is intractable even for small networks due to the large state–action space, making classical dynamic programming impractical. To address this challenge, we develop an Approximate Dynamic Programming (ADP) approach \cite{wang2019markov, wang2020mdp} that leverages post-decision states to simplify the Bellman recursion. A post-decision state represents the system immediately after a switching action and before uncertainties resolve, allowing the expectation over future conditions to be separated from the optimization step. Through iterative simulation of wildfire scenarios and value-function approximation, the ADP algorithm learns the long-term value of post-decision states and yields a near-optimal switching policy for PSPS operations. This approach enables a tractable solution of the sequential switching decisions without enumerating all possible state trajectories.

We evaluate the proposed framework on 54-bus and 138-bus distribution test systems. The results demonstrate that the optimized switching policy substantially reduces total wildfire-related costs compared with baseline strategies based on static, hour-by-hour deterministic optimization. The learned policy proactively de-energizes lines carrying high current when wildfire risk is elevated, while avoiding unnecessary outages by restoring service promptly once local conditions improve. This allows the operator to maintain reliability without exposing the system to hazardous configurations.  Overall, the proposed MDP-based framework and ADP solution provide an effective tool for utilities seeking to balance safety and reliability during wildfire events.
\vspace{-0.3cm}
\section{Model Formulation}\label{sec:model}
\vspace{-0.2cm}
In this section, we present our distributionally robust Markov Decision Process (DRMDP) formulation for wildfire-aware grid reconfiguration optimization. The model captures sequential switching decisions under endogenous and exogenous uncertainty, focusing on fire-prone infrastructure within distribution networks. We present the different MDP components as follows:

\paragraph{\bf States} The system state, denoted by $\boldsymbol{s}$, characterizes the operational condition of the distribution network and is detailed as \textcolor{black}{$\boldsymbol{s}=\big[\boldsymbol{a}^{avail^{\top}},\,\boldsymbol{f}^{r^{\top}},\,\boldsymbol{z}^{\mathrm{sw},0^{\top}},\,\boldsymbol{D}^{p^{\top}},\,\boldsymbol{D}^{q^{\top}}\big]^{\top}$}. Vectors $\boldsymbol{a}^{avail}=[a^{avail}_l]_{l\in\mathcal{L}}$ and $\boldsymbol{f}^{fire}=[f^{fire}_l]_{l\in\mathcal{L}}$ describe the condition of each line $l \in \mathcal{L}$: $a^{avail}_l\in\{0,1\}$ indicates whether line $l$ is available for service or failed, and $f^{fire}_l\in\{0,1\}$ flags whether line $l$ is currently fire-affected. The vector $\boldsymbol{z}^{\mathrm{sw},0}=[z^{\mathrm{sw},0}_l]_{l\in\mathcal{L}^{\mathrm{sw}}}$ records the current pre-decision switching status of switchable lines $\mathcal{L}^{\mathrm{sw}}\subseteq \mathcal{L}$, where $z^{\mathrm{sw},0}_l=1$ denotes closed and $z^{\mathrm{sw},0}_l=0$ denotes open. Finally, $\boldsymbol{D}^p=[D^p_b]_{b\in\mathcal{N}}$ and $\boldsymbol{D}^q=[D^q_b]_{b\in\mathcal{N}}$ are the active and reactive demands at buses in the node set $\mathcal{N}$.

\paragraph{\bf Actions}
At each stage, the operator chooses a switching configuration and an operating point for the resulting network topology. We denote the action by 
$\boldsymbol{a}=\big[\boldsymbol{y}^{\mathrm{sw}^{\top}},\,\boldsymbol{z}^{\mathrm{sw}^{\top}},\,\boldsymbol{d}^{\top},\,\boldsymbol{w}^{\mathrm{op}^{\top}},\,\boldsymbol{f}^{p^{\top}},\,|\boldsymbol{f}^{p^{\top}}|,\,\boldsymbol{f}^{q^{\top}},\,\boldsymbol{v}^{\top},\,\boldsymbol{p}^{\text{sub}^{\top}},\,\boldsymbol{q}^{\text{sub}^{\top}},\\\,\boldsymbol{\Delta D}^{p\pm^{\top}},\,\boldsymbol{\Delta D}^{q\pm^{\top}},\,\boldsymbol{\iota}^{\top}\big]^{\top}$. For each switchable line $l\in\mathcal{L}^{\mathrm{sw}}$, $y^{\mathrm{sw}}_l\in\{0,1\}$ is the switching operation (1 if line $l$ is switched on, 0 if switched off), and $z^{\mathrm{sw}}_l\in\{0,1\}$ is the resulting line status after applying the switching action. For each line $l\in\mathcal{L}$, $d_l\in\{0,1\}$ encodes the chosen reference direction for flow on line $l$, and $w^\mathrm{op}_l$ indicates whether the line is operational in the current topology. Finally, for each bus $b$, $\iota_b$ is a binary indicator that equals 1 if bus $b$ is electrically isolated from the power network and 0 otherwise.

The operator also chooses continuous operating variables: $f^{p}_l$ and $f^q_l$ are the active and reactive line flows, represented using a split formulation $f^p_l=f^{p+}_l-f^{p-}_l$ with $f^{p+}_l\ge 0$ and $f^{p-}_l\ge 0$. Let $\boldsymbol{v}=[v_b]_{b\in\mathcal{N}}$ denote squared voltage magnitudes at buses. At substation buses $\mathcal{N}^{\mathrm{sub}}\subseteq\mathcal{N}$, $p_b^{\mathrm{sub}}$ and $q_b^{\mathrm{sub}}$ denote active and reactive power injections. Finally, load-balance slack variables $\Delta D^{p+}_b,\Delta D^{p-}_b,\Delta D^{q+}_b$, and $\Delta D^{q-}_b$ quantify active/reactive demand shortfall (load shedding) and surplus (over-supply) at each bus $b\in\mathcal{N}$.

\paragraph{\bf Transition Probabilities}
State transitions are driven by the evolution of line availability $a^{avail}_l$. Following \cite{moreira2024distribution}, we model the next-period availability of each line $l$ as a Bernoulli random variable whose success probability depends on both wildfire exposure and endogenous power flow. Let $\gamma_l\in[0,1]$ be the baseline probability that line $l$ remains operational over one period in the absence of wildfire, and let $\beta_l\ge 0$ quantify how higher active power flow increases failure risk. We use $f^{fire}_{lt}\in\{0,1\}$ to indicate whether line $l$ is wildfire-affected at time $t$, and denote the realized active power flow on line $l$ at time $t$ by $f_{lt}^{p}$.

For a line that is currently available ($a^{avail}_{lt}=1$), the next-period availability depends on whether the line is in the wildfire zone and whether fire has reached the line. Let $\mathcal{L}^{fr}\subseteq\mathcal{L}$ denote the set of lines in the wildfire zone. For $l\in\mathcal{L}^{fr}$, the indicator $f^{fire}_{lt}\in\{0,1\}$ denotes whether fire has reached line $l$ at time $t$. The next-period availability is
\begin{align}
    &\mathbb{P}\!\left(a^{avail}_{l,t+1}=1 \,\middle|\, a^{avail}_{l,t} = 1,\, f^{fire}_{l,t},\, |f_{l,t}^{p}|\right) \notag \\
    &\hspace{50pt} =\begin{cases}
        (1-f^{fire}_{l,t})\big(\gamma_l-\beta_l \left|f_{l,t}^{p}\right|\big), & l\in\mathcal{L}^{fr},\\[2pt]
        1, & l\notin\mathcal{L}^{fr}.
    \end{cases}
    \label{eq:tran_prob}
\end{align}
This specification has two cases for wildfire-zone lines. If fire has reached the line with $f^{fire}_{lt}=1$, we set the available probability to zero for the next period. If the line is in the wildfire zone but not yet reached by fire with $f^{fire}_{lt}=0$, it remains available with probability $\gamma_l-\beta_l|f_{lt}^{p}|$, which decreases with active power flow. For lines outside the wildfire zone $l\notin\mathcal{L}^{fr}$, we assume no wildfire-induced outage during the period, hence $\mathbb{P}(a^{avail}_{l,t+1}=1\mid a^{avail}_{lt}=1)=1$. Once a line has failed ($a^{avail}_{lt}=0$), it remains unavailable for the remainder of the horizon, i.e., $\mathbb{P}(a^{avail}_{l,t+1}=1\mid a^{avail}_{lt}=0)=0$.

Because $f_{lt}^{p}$ belongs to the stage-$t$ action, \eqref{eq:tran_prob} couples switching and dispatch decisions to future network availability, which is the source of decision-dependent uncertainty in the DRMDP.

\paragraph{\bf Rewards}
We use a stage reward equal to the negative operating cost of the distribution network incurred at time $t$. The cost has three parts: (i) the cost of active power purchased from substations, (ii) the cost of intentional switching actions, and (iii) penalties on unsupplied or excess active and reactive power, which act as surrogates for load shedding, over-supply, and poor voltage support. Formally, the reward at time $t$ is:
\begin{align}
    r_{\boldsymbol{as},t} &= 
    -\sum_{b \in \mathcal{N}^{\text{sub}}} C^{\text{energy}} \cdot p^{\text{sub}}_{b,t} - \sum_{l \in \mathcal{L}^{\text{sw}}} C^{\text{switch}} \cdot y_{l,t}^{\mathrm{sw}} \notag \\
    & \hspace{11pt}  - \sum_{b \in \mathcal{N}^{\text{all}}} C^{\text{load\_loss}} \cdot \Bigl ( \Delta D^{p+}_{b,t} + \Delta D^{p-}_{b,t}\notag \\
    & \hspace{11pt}  + \Delta D^{q+}_{b,t} + \Delta D^{q-}_{b,t} \Bigr ) + \alpha.
\end{align}
Here, $C^{\text{energy}}$ is the unit energy cost, $C^{\text{switch}}$ is the per-operation switching penalty, and $C^{\text{load\_loss}}$ captures the cost of active and reactive power imbalances at each bus. The term $\alpha$ represents the decision-dependent expected second-stage cost. This reward formulation promotes economically efficient operation while discouraging excessive switching and large amounts of unmet demand. It is integrated into the DRMDP Bellman recursion to guide resilient switching decisions under uncertainty.

\vspace{-0.5cm}
\subsection{Distributionally Robust Bellman Equation}
\vspace{-0.2cm}

As discussed above, the operator’s objective is to minimize the cumulative operational cost under the worst-case distribution of future state transitions. The distributionally robust value function is defined recursively as:
\begin{align}
    & V_t(\boldsymbol{s}) = \max_{\boldsymbol{a}\in\mathcal{A}(\boldsymbol{s})} \min_{\mu_{\boldsymbol{as}} \in \mathcal{D}_{\boldsymbol{as}}} \mathbb{E}_{\boldsymbol{p}_{\boldsymbol{as}}\sim \mu_{\boldsymbol{as}}} \left[r_{\boldsymbol{as},t} + \lambda \boldsymbol{p}_{\boldsymbol{as}}^T \boldsymbol{V}_{t+1}\right] \notag\ \\
    &\hspace{10pt} = \max_{\boldsymbol{a}\in\mathcal{A}(\boldsymbol{s})} \min_{\mu_{\boldsymbol{as}} \in \mathcal{D}_{\boldsymbol{as}}}\left[ r_{\boldsymbol{as},t} + \int_{\boldsymbol{p}}\lambda \boldsymbol{p}_{\boldsymbol{as}}^T\boldsymbol{V}_{t+1}d\mu(\boldsymbol{p}_{\boldsymbol{as}})\right] \notag \\
    &\hspace{10pt} = \max_{\boldsymbol{a}\in\mathcal{A}(\boldsymbol{s})} ~ \left[r_{\boldsymbol{as},t} + \min_{\mu_{\boldsymbol{as}} \in \mathcal{D}_{\boldsymbol{as}}}  \int_{\boldsymbol{p}}\lambda \boldsymbol{p}_{\boldsymbol{as}}^T\boldsymbol{V}_{t+1}d\mu(\boldsymbol{p}_{\boldsymbol{as}})\right] .
\end{align}
where $V_{t}(\boldsymbol{s})$ is the optimal value at time $t$, $r_{\boldsymbol{as},t}$ is the immediate reward of taking action $\boldsymbol{a}$, and $\lambda\in(0,1)$ is the discount factor. The vector $\boldsymbol{V}_{t+1}$ stacks the next-stage values $\{V_{t+1}(\boldsymbol{s}')\}_{\boldsymbol{s}'\in\mathcal{S}}$, and $\boldsymbol{p}_{\boldsymbol{a}\boldsymbol{s}}$ is the corresponding transition-probability vector over next states. The transition law is uncertain: $\mu_{\boldsymbol{a}\boldsymbol{s}}$ is a distribution over $\boldsymbol{p}_{\boldsymbol{a}\boldsymbol{s}}$ and is chosen from the ambiguity set $\mathcal{D}_{\boldsymbol{a}\boldsymbol{s}}$.

This recursion induces a robust state-action value function. For any $(\boldsymbol{s},\boldsymbol{a})$, define
\begin{align}
Q_{\boldsymbol{as},t} = r_{\boldsymbol{as},t} + \min_{\mu_{\boldsymbol{as}} \in \mathcal{D}_{\boldsymbol{as}}} \int\lambda \boldsymbol{p}_{\boldsymbol{as}}^T\boldsymbol{V}_{t+1}d\mu(\boldsymbol{p}_{\boldsymbol{as}}),
\end{align}
so that
\begin{align}
V_{t}(\boldsymbol{s})=\max_{\boldsymbol{a}\in\mathcal{A}(\boldsymbol{s})} Q_{t}(\boldsymbol{s},\boldsymbol{a}).
\end{align}

\vspace{-0.5cm}
\subsection{Endogenous Ambiguity Set}\label{sec:DDU}
We define a distributionally robust ambiguity set $\mathcal{D}_{\boldsymbol{as}}$ to capture uncertainty in the transition probabilities $\boldsymbol{p}_{\boldsymbol{as}}$, which are endogenously affected by the operator’s actions. Switching decisions influence line flows and, consequently, the likelihood of line failure, making the transition probabilities decision-dependent. Let $\{p_{\boldsymbol{as},i}(\cdot)\}_{i=1}^{N}$ denote a finite set of candidate transition distributions. The ambiguity set is then formulated as:
\begin{align}
    &\mathcal{D}_{\boldsymbol{as}}:= \bigg\{ \sum_{i=1}^N q_i p_{\boldsymbol{as},i}(\boldsymbol{s}')\ \bigg | \
 q_i \ge 0, \ \sum_{i=1}^N q_i = 1, \notag \\
    & \hspace{180pt} \ \forall \boldsymbol{s}'\in \mathcal{S} \bigg\},
\end{align}
where $p_{\boldsymbol{as},i}(\boldsymbol{s}')$ are decision-dependent transition probability to successor state $\boldsymbol{s}' \in \mathcal{S}$, and the mixture weights $\boldsymbol{q}$ parameterize an adversarially chosen convex combination within $\mathcal{D}_{\boldsymbol{a}\boldsymbol{s}}$.

Using this discrete mixture representation, the distributionally robust value function admits a standard reformulation. Substituting $\mathcal{D}_{\boldsymbol{as}}$ into the Bellman recursion yields:
\begin{align}
    & V_t(\boldsymbol{s}) = \max_{\boldsymbol{a}\in\mathcal{A}(\boldsymbol{s})} ~ \Biggl [r_{\boldsymbol{as},t} + \min_{\substack{q_i \ge 0 \\ \sum_{i=1}^N q_i = 1}}  \notag\\
    &\hspace{76pt} \lambda \sum_{s'\in\mathcal{S}}\sum_{i=1}^N q_i p_{\boldsymbol{as},i}(\boldsymbol{s}') V_{t+1}(\boldsymbol{s}') \Biggr ]
    \label{eq:dr_bellman_mixture}
\end{align}
The action $\boldsymbol{a}$ enters \eqref{eq:dr_bellman_mixture} both through the immediate reward and through the candidate transition models $p_{\boldsymbol{a}\boldsymbol{s},i}(\cdot)$, reflecting decision-dependent uncertainty. The discrete-mixture form makes the worst-case transition selection explicit and yields a tractable representation for computing the DRMDP.

\subsection{Reformulation via Primal-Dual Optimization}
At time $t$ in state $\boldsymbol{s}$, the operator selects an action $\boldsymbol{a}\in\mathcal{A}(\boldsymbol{s})$. The transition law is then chosen adversarially from the discrete-mixture ambiguity set, which amounts to selecting mixture weights $\boldsymbol{q}=(q_1,\dots,q_N)$ over $N$ candidate transition models. For each candidate model $i=1,\dots,N$, define the discounted continuation value
\begin{equation}
    g_{i,t+1}(\boldsymbol{s},\boldsymbol{a})
    :=\lambda\sum_{\boldsymbol{s}'\in\mathcal{S}} p_{\boldsymbol{a}\boldsymbol{s},i}(\boldsymbol{s}')\,V_{t+1}(\boldsymbol{s}').
\label{eq:g_def}
\end{equation}
Then the inner minimization is the linear program
\begin{align}
    \min_{\boldsymbol{q}}\quad & \sum_{i=1}^{N} q_i\, g_{i,t+1}(\boldsymbol{s},\boldsymbol{a}) \label{eq:primal_q_simple}\\
    \text{s.t.}\quad & \sum_{i=1}^{N} q_i = 1 \ :(\alpha), \\
    &q_i\ge 0 \ :(\pi_i),\ \ i=1,\dots,N. 
\end{align}
This problem is convex in $\boldsymbol{q}$. Introducing the Lagrange multiplier $\alpha\in\mathbb{R}$ for the equality constraint and $\pi_i\ge 0$ for $q_i\ge 0$, its dual can be written in the equivalent, simplified form
\begin{align}
    \max_{\alpha\in\mathbb{R}}\quad & \alpha \label{ob:dual_q_refined}\\
    \text{s.t.}\quad &
    \alpha \le g_{i,t+1}(\boldsymbol{s},\boldsymbol{a}),\qquad i=1,\dots,N,
    \label{eq:dual_q_refined}
\end{align}
where the optimal value satisfies
\begin{equation}
    \alpha^\star=\min_{i=1,\dots,N} g_{i,t+1}(\boldsymbol{s},\boldsymbol{a})
\end{equation}

Substituting this dual representation back into the Bellman recursion yields 
\begin{align}
    & V_{t}(\boldsymbol{s})
    =\max_{\boldsymbol{a}\in\mathcal{A}(\boldsymbol{s})}
    \Bigl \{
    r_{\boldsymbol{a}\boldsymbol{s},t}+\max_{\alpha\in\mathbb{R}}
    \left\{\alpha:\ \alpha\le g_{i,t+1}(\boldsymbol{s},\boldsymbol{a}),\ \forall i\right\}
    \Bigr \} \notag\\
    & \hspace{10pt} =\max_{\boldsymbol{a}\in\mathcal{A}(\boldsymbol{s})}
    \Bigl \{
    r_{\boldsymbol{a}\boldsymbol{s},t}
    +\min_{i=1,\dots,N} g_{i,t+1}(\boldsymbol{s},\boldsymbol{a})
    \Bigr \} \notag\\
    & \hspace{10pt} =\max_{\boldsymbol{a}\in\mathcal{A}(\boldsymbol{s})}
    \Bigl \{
    r_{\boldsymbol{a}\boldsymbol{s},t}
    \notag \\
    & \hspace{65pt} +\min_{i=1,\dots,N}\;
    \lambda\sum_{\boldsymbol{s}'\in\mathcal{S}} p_{\boldsymbol{a}\boldsymbol{s},i}(\boldsymbol{s}')\,V_{t+1}(\boldsymbol{s}')
    \Bigr \}.
    \label{eq:dr_bellman_min_i_refined}
\end{align}
Equation \eqref{eq:dr_bellman_min_i_refined} shows that, for a fixed action, the worst-case mixture over candidate transition models is attained at an extreme point: nature selects the candidate transition model that yields the smallest expected continuation value.

\section{Implementation of the Transition Probability}\label{sec:trans_prob}

This section describes the approximation of the decision-dependent line-availability transition probabilities used within the DRMDP framework. To capture the uncertainty in line failures and their dependence on operational decisions, we present both the basic probabilistic formulation and its linearized approximation for computational implementation. In particular, as shown in \eqref{eq:dual_q_refined}, the term $g_{i,t+1}(\boldsymbol{s},\boldsymbol{a})$ must be expressed as a linear function of the action $\boldsymbol{a}$ to ensure the optimization model \eqref{ob:dual_q_refined}-\eqref{eq:dual_q_refined} remains tractable for off-the-shelf solvers. This requirement, in turn, motivates approximating each candidate transition probability $p_{\boldsymbol{a}\boldsymbol{s},i}$ by a form that is linear in $\boldsymbol{a}$, thereby enabling efficient evaluation of the robust Bellman recursion.
\vspace{-0.4cm}
\subsection{Basic Formulation} \label{sec:basic}
We assume that line failures occur independently across the network and are driven by local operating conditions and wildfire simulation. For any line located in the wildfire zone $l \in \mathcal{L}^{fr}$ that is available at time $t$ (i.e., $a^{avail}_{lt} = 1$), the probability that it remains available at time $t+1$ is modeled as:
\begin{align}
    & P(a^{avail}_{l,t+1} = 1 | a^{avail}_{lt} = 1, f^{fire}_{lt}) \notag \\
    & \hspace{100pt} = (1 - f^{fire}_{lt})(\gamma_l - \beta_l |f_{lt}^{p}|).
\end{align}
In practice, we assume that $\gamma$ and $\beta$ are homogeneous across all lines. The factor $(1 - f^{fire}_{lt})$ acts as a wildfire-simulation mask and is omitted in subsequent expressions for simplicity, as it is externally known.

Given this per-line failure model, we now define the transition probability from a current state $s$ to a successor state $s'$ under action $\boldsymbol{a}$. Because a failed line cannot return to service within the short decision horizon, transitions of the type $0 \rightarrow 1$ are excluded from the feasible state set $\mathcal{S}$. For all feasible transitions, the probability of reaching state $s'$ is
\begin{equation}
    p_{\boldsymbol{as}}(\boldsymbol{s}')
    = 1^{n_{00}^{\boldsymbol{s}'}}\prod_{l\in\mathcal{L}_{11}^{\boldsymbol{s}'}}(\gamma_l-\beta_l \left|f_{lt}^{p}\right|)\prod_{l\in\mathcal{L}_{10}^{\boldsymbol{s}'}}(1-\gamma_l+\beta_l \left|f_{lt}^{p}\right|),
\end{equation}
where
\begin{itemize}
    \item $\mathcal{L}_{11}^{\boldsymbol{s}'}$ denotes the set of lines that remain available ($1 \rightarrow 1$),
    \item $\mathcal{L}_{10}^{\boldsymbol{s}'}$ denotes the set of lines that fail during the transition ($1 \rightarrow 0$),
    \item $n_{00}^{\boldsymbol{s}'}$ denotes the number of lines that remain unavailable ($0 \rightarrow 0$),
    \item Transitions $0 \rightarrow 1$ are considered infeasible and excluded from $\mathcal{S}$.
\end{itemize}
\vspace{-0.4cm}
\subsection{Linear Approximation}

To enable efficient computation and integration with optimization solutions, we develop a first-order linear approximation of the transition probability $p_{\boldsymbol{as}}(\boldsymbol{s}')$ around the nominal point $\boldsymbol{a}=\boldsymbol{0}$. The only element within the action vector $\boldsymbol{a}$ that corresponds to the absolute value of the active power flow through line $l$ is $\left|f_{lt}^{p}\right|$. 

Dropping the constant factor $1^{n_{00}^{\boldsymbol{s}'}}$, define
\begin{equation}
    F(\boldsymbol{a}) = \prod_{l\in\mathcal{L}_{11}^{\boldsymbol{s}'}}(\gamma_l-\beta_l \left|f_{lt}^{p}\right|)\prod_{l\in\mathcal{L}_{10}^{\boldsymbol{s}'}}(1-\gamma_l+\beta_l \left|f_{lt}^{p}\right|).
\end{equation}
Evaluating at $\boldsymbol{a}=\boldsymbol{0}$ gives the baseline transition probability
\begin{equation}
    F(\boldsymbol{0}) = \prod_{l\in\mathcal{L}_{11}^{\boldsymbol{s}'}}\gamma_l\prod_{l\in\mathcal{L}_{10}^{\boldsymbol{s}'}}(1-\gamma_l).
\end{equation}
We next compute the partial derivatives of $F(\boldsymbol{a})$ with respect to each $\left|f_{lt}^{p}\right|$.  
Three cases arise:
\begin{itemize}
    \item If $l \notin \mathcal{L}_{10}^{\boldsymbol{s}'}\cup \mathcal{L}_{11}^{\boldsymbol{s}'}$, i.e., line $l$ goes $0\rightarrow0$, then $\frac{\partial F}{\partial \left|f_{lt}^{p}\right|}\equiv 0$.
    \item If $l \in \mathcal{L}_{11}^{\boldsymbol{s}'}$, the factor $(\gamma_l-\beta_l \left|f_{lt}^{p}\right|)$ appears once in the product, and we can obtain:
    \begin{align}
        & \frac{\partial F}{\partial \left|f_{lt}^{p}\right|}(\boldsymbol{a})=-\beta_l\prod_{l'\in\mathcal{L}_{11}^{\boldsymbol{s}'}\setminus \{l\}}(\gamma_{l'}-\beta_{l'} \left|f_{l',t}^{p}\right|) \notag \\ 
        & \hspace{78pt} \prod_{l''\in\mathcal{L}_{10}^{\boldsymbol{s}'}}(1-\gamma_{l''}+\beta_{l''} \left|f_{l'',t}^{p}\right|).
    \end{align}
    Evaluating at $\boldsymbol{a}=\boldsymbol{0}$,
    \begin{equation}
        \frac{\partial F}{\partial \left|f_{lt}^{p}\right|}\bigg|_{\boldsymbol{0}}=- \beta_l \prod_{l'\in\mathcal{L}_{11}^{\boldsymbol{s}'}\setminus \{l\}} \gamma_{l'} \prod_{l''\in\mathcal{L}_{10}^{\boldsymbol{s}'}} (1-\gamma_{l''}).
    \end{equation}
    \item If $l \in \mathcal{L}_{10}^{\boldsymbol{s}'}$, the factor $(1-\gamma_l+\beta \left|f_{lt}^{p}\right|)$ appears once in the product, and we can get:
    \begin{align}
        &\frac{\partial F}{\partial \left|f_{lt}^{p}\right|}(\boldsymbol{a})=\beta_l\prod_{l'\in\mathcal{L}_{11}^{\boldsymbol{s}'}}(\gamma_{l'}-\beta_{l'} \left|f_{l',t}^{p}\right|) \notag \\
        &\hspace{65pt} \prod_{l''\in\mathcal{L}_{10}^{\boldsymbol{s}'}\setminus \{l\}}(1-\gamma_{l''}+\beta_{l''} \left|f_{l'',t}^{p}\right|).
    \end{align}
    Evaluating at $\boldsymbol{a}=\boldsymbol{0}$,
    \begin{equation}
        \frac{\partial F}{\partial \left|f_{lt}^{p}\right|}\bigg|_{\boldsymbol{0}}= \beta_l\prod_{l'\in\mathcal{L}_{11}^{\boldsymbol{s}'}}\gamma_{l'}\prod_{l''\in\mathcal{L}_{10}^{\boldsymbol{s}'}\setminus \{l\}}(1-\gamma_{l''}).
    \end{equation}
\end{itemize}
Combining the above, the first-order Taylor approximation of $p_{\boldsymbol{as}}(\boldsymbol{s}') $ around $\boldsymbol{0}$ is
\begin{align}
    &p_{\boldsymbol{as}}(\boldsymbol{s}') \approx p_{\boldsymbol{0s}}(\boldsymbol{s}')+\sum_{l\in\mathcal{L}} \frac{\partial F}{\partial \left|f_{lt}^{p}\right|}\bigg|_{\boldsymbol{0}}\left|f_{lt}^{p}\right|\\ \nonumber
    &= \prod_{l\in\mathcal{L}_{11}^{\boldsymbol{s}'}}\gamma_l\prod_{l\in\mathcal{L}_{10}^{\boldsymbol{s}'}}(1-\gamma_l)\\ \nonumber
    &-\sum_{l\in\mathcal{L}_{11}^{\boldsymbol{s}'}}\beta_l \prod_{l'\in\mathcal{L}_{11}^{\boldsymbol{s}'}\setminus \{l\}} \gamma_{l'} \prod_{l''\in\mathcal{L}_{10}^{\boldsymbol{s}'}} (1-\gamma_{l''})\left|f_{lt}^{p}\right| \\ \nonumber
    &+ \sum_{l\in\mathcal{L}_{10}^{\boldsymbol{s}'}}\beta_l\prod_{l'\in\mathcal{L}_{11}^{\boldsymbol{s}'}}\gamma_{l'}\prod_{l''\in\mathcal{L}_{10}^{\boldsymbol{s}'}\setminus \{l\}}(1-\gamma_{l''})\left|f_{lt}^{p}\right|. \nonumber
\end{align}

Although the exact transition probabilities satisfy $\sum_{\boldsymbol{s}'} p_{\boldsymbol{as}}(\boldsymbol{s}') = 1$, the linearized approximation may not.  
Therefore, we apply a normalization step:
\begin{equation}
    \tilde{p}_{\boldsymbol{as}}(\boldsymbol{s}') = \frac{p_{\boldsymbol{as}}(\boldsymbol{s}')}{\sum_{s''}\hat{p}_{\boldsymbol{as}}(\boldsymbol{s}'')}.
\end{equation}
The normalized linear model $\widetilde{p}_{\boldsymbol{as}}(\boldsymbol{s}')$ serves as a computationally efficient approximation of the true transition probabilities and is compatible with the DRMDP solution framework.

\subsection{Candidate Distributions}
\vspace{-0.2cm}
As described in Section~\ref{sec:DDU}, we consider a finite set of $N$ candidate transition models $p_{\boldsymbol{a}\boldsymbol{s},i}(\cdot)$.  
These candidate models capture uncertainty in the parameters $(\gamma_l,\beta_l)$ that govern line‐failure probabilities.  
For each candidate $i$, we specify parameters $(\gamma_{li},\beta_{li})$ and define the exact (nonlinear) transition probability by
\vspace{-0.2cm}
\begin{equation}
    p_{\boldsymbol{a}\boldsymbol{s},i}(\boldsymbol{s}')=\prod_{l\in\mathcal{L}_{11}^{\boldsymbol{s}'}}(\gamma_{li}-\beta_{li} \left|f_{lt}^{p}\right|)\prod_{l\in\mathcal{L}_{10}^{\boldsymbol{s}'}}(1-\gamma_{li}+\beta_{li} \left|f_{lt}^{p}\right|).
\end{equation}
To obtain a computationally efficient approximation, we linearize $p_i(s' | s,\boldsymbol{a})$ around $\boldsymbol{a}=\boldsymbol{0}$.  
The resulting affine form is
\vspace{-0.2cm}
\begin{equation}
    p_{\boldsymbol{a}\boldsymbol{s},i}(\boldsymbol{s}')\approx c_{0,i}(\boldsymbol{s}') + \sum_{l\in\mathcal{L}}c_{l,i}(\boldsymbol{s}')\left|f_{lt}^{p}\right|,
\end{equation}
where
\vspace{-0.2cm}
\begin{align}
    c_{0,i}(\boldsymbol{s}') = p_{\boldsymbol{0}\boldsymbol{s},i}(\boldsymbol{s}') =\prod_{l\in\mathcal{L}_{11}^{\boldsymbol{s}'}}\gamma_{li}\prod_{l\in\mathcal{L}_{10}^{\boldsymbol{s}'}}(1-\gamma_{li}), 
\end{align}
and the coefficients
\vspace{-0.2cm}
\begin{align}
    &c_{l,i}(\boldsymbol{s}') =\begin{cases}
        -\beta_{li} \prod_{l'\in\mathcal{L}_{11}^{\boldsymbol{s}'}\setminus \{l\}} \gamma_{l',i} \prod_{l''\in\mathcal{L}_{10}^{\boldsymbol{s}'}} (1-\gamma_{l'',i}), %
        \notag\\
        \hspace{130pt} \text{if } l\in\mathcal{L}_{11}^{\boldsymbol{s}'}\\
\beta_{li}\prod_{l'\in\mathcal{L}_{11}^{\boldsymbol{s}'}}\gamma_{l',i}\prod_{l''\in\mathcal{L}_{10}^{\boldsymbol{s}'}\setminus \{l\}}(1-\gamma_{l'',i}), 
\notag\\
        \hspace{130pt} \text{if } l \in\mathcal{L}_{10}^{\boldsymbol{s}'}\\
        0, \text{Otherwise}
    \end{cases}
\end{align}
\vspace{-0.2cm}
\begin{align}
    c_{li}(\boldsymbol{s}')
    =
    \begin{cases}
        -\beta_{li} \,(1-\gamma_{li})\,\bar{\Phi}_{-l,i}(\boldsymbol{s}'),
        & \text{if }l \in \mathcal{L}_{11}^{\boldsymbol{s}'} \\[6pt]
        \beta_{li} \,\gamma_{li}\,\bar{\Phi}_{-l,i}(\boldsymbol{s}'),
        & \text{if }l \in \mathcal{L}_{10}^{\boldsymbol{s}'} \\[6pt]
        0,
        & \text{Otherwise}
    \end{cases}
\end{align}
where
\vspace{-0.2cm}
\begin{align}
    \bar{\Phi}_{-l,i}(\boldsymbol{s}')
    :=
    \prod_{l'\in \mathcal{L}_{11}^{\boldsymbol{s}'} \setminus \{l\}}
    \gamma_{l',i}
    \;
    \prod_{l''\in \mathcal{L}_{10}^{\boldsymbol{s}'} \setminus \{l\}}
    (1-\gamma_{l'',i}).
\end{align}

Thus, the linear approximation for candidate $i$ takes the form
\begin{align}
    & \hat{p}_{\boldsymbol{a}\boldsymbol{s},i}(\boldsymbol{s}') = \gamma_{li}(1-\gamma_{li})\,\bar{\Phi}_{-l,i}(\boldsymbol{s}') \notag \\ 
    & \hspace{60pt} -\sum_{l\in\mathcal{L}_{11}^{\boldsymbol{s}'}}\beta_{li} \,(1-\gamma_{li})\,\bar{\Phi}_{-l,i}(\boldsymbol{s}')\left|f_{lt}^{p}\right| \notag \\ 
    & \hspace{60pt} + \sum_{l\in\mathcal{L}_{10}^{\boldsymbol{s}'}}\beta_{li} \,\gamma_{li}\,\bar{\Phi}_{-l,i}(\boldsymbol{s}')\left|f_{lt}^{p}\right|.
\end{align}
Because this linear approximation does not necessarily satisfy the probability normalization condition  $\sum_{\boldsymbol{s}'}\tilde{p}_{\boldsymbol{a}\boldsymbol{s},i}(\boldsymbol{s}')=1$, we apply the following normalization:
\begin{equation}
    \tilde{p}_{\boldsymbol{a}\boldsymbol{s},i}(\boldsymbol{s}') = \frac{\hat{p}_{\boldsymbol{a}\boldsymbol{s},i}(\boldsymbol{s}')}{\sum_{s''\in\mathcal{S}}\hat{p}_{\boldsymbol{a}\boldsymbol{s},i}(\boldsymbol{s}'')}.
\end{equation}



\section{Distribution System Operation}\label{sec:DSO}

This section presents the optimization framework for operating wildfire-prone distribution systems under uncertainty. The operator seeks to minimize the worst-case expected operational cost over a finite horizon, considering line availability, switching actions, and network constraints such as voltage limits and power flow physics. The decision process is modeled as a distributionally robust MDP, and computational tractability is achieved through an Approximate Dynamic Programming (ADP) method. The ADP algorithm estimates long-term values using post-decision states and updates these estimates through simulation and regression, enabling adaptive and resilient switching policies under uncertain line failures.
\subsection{Recursive Optimization Objective}
Let $V_{t}(\boldsymbol{s})$ denote the value function at state $s \in \mathcal{S}$ and time $t$.  
The operator aims to minimize the total cost of system operation, which includes the active power procurement, switching cost, and penalties for load imbalance and unmet demand (both active and reactive).  Under the distributionally robust framework, the one-step reward plus value-to-go can be written as:
\begin{align}
    & V_t(\boldsymbol{s})= \underset{{\substack{ p^{\text{sub}}_{bt}, y^{sw}_l, \Delta D^{p+}_b,\\ \Delta D^{p-}_b, \Delta D^{q+}_b, \Delta D^{q-}_b }}}{\text{Maximize}} \hspace{-0.1cm} -\sum_{b \in N_{\text{sub}}} C^{\text{energy}} \cdot p^{\text{sub}}_{bt} \notag \\
    & \hspace{10pt} - \sum_{l \in \mathcal{L}^{\text{sw}}} C^{\text{switch}} \cdot y_{lt}^{\mathrm{sw}} \label{eq:recur_obj} - \sum_{b \in \mathcal{N}^{\text{all}}} C^{\text{load\_loss}} \cdot \Biggl ( \Delta D^{p+}_{bt} \notag \\
    & \hspace{70pt}+ \Delta D^{p-}_{bt} + \Delta D^{q+}_{bt} + \Delta D^{q-}_{bt} \Biggr )  + \alpha  \\ 
    & \text{s.t.} \notag \\ 
    & \alpha \le \lambda \sum_{\boldsymbol{s}'\in \mathcal{S}}p_{\boldsymbol{as},i}(\boldsymbol{s}')V_{t+1}(\boldsymbol{s}'), \ \ \forall i = 1,\dots, N, \label{eq:dual_constraint}
\end{align}
    
\noindent where $\alpha$ is the dual variable arising from the reformulation of the inner minimization in the distributionally robust Bellman operator.

Since each transition probability $p_{\boldsymbol{as},i}(\boldsymbol{s}')$ is approximated linearly as $c_{0,i}(\boldsymbol{s}') + \sum_{l\in\mathcal{L}}c_{l,i}(\boldsymbol{s}')\left|f_{lt}^{p}\right|$, substituting this into the robust constraint~\eqref{eq:dual_constraint} yields
\begin{align}
    & \alpha\le \lambda \sum_{\boldsymbol{s}'\in \mathcal{S}} \Bigl [c_{0,i}(\boldsymbol{s}')V_{t+1}(\boldsymbol{s}') \notag \\
    & \hspace{17pt} +\sum_{l\in\mathcal{L}}c_{l,i}(\boldsymbol{s}')\left|f_{lt}^{p}\right| V_{t+1}(\boldsymbol{s}') \Bigr ], \forall i = 1,\dots, N.
\end{align}

\subsection{Operational Constraints}\label{sec:oper}
The distribution system operation is governed by power balance equations, voltage limits, thermal limits, and switching feasibility rules.  
We summarize the complete constraint set below. 

For each substation $b\in \mathcal{N}^{\text{subs}}$:
\begin{align}
    & p^{\text{sub}}_{bt} + \sum_{l\in\mathcal{L}|to(l)=b} \left(f^{p+}_{lt} - f^{p-}_{lt}\right) \notag \\
    & \hspace{10pt} - \sum_{l\in\mathcal{L}|fr(l)=b} \left(f^{p+}_{lt} - f^{p-}_{lt}\right)- D^{p}_{bt} - \Delta D^{p+}_{bt} \notag \\
    & \hspace{130pt} + \Delta D^{p-}_{bt} = 0 \\
    & q^{\text{sub}}_{bt} + \sum_{l\in\mathcal{L}|to(l)=b} f^{q}_{lt} - \sum_{l\in\mathcal{L}|fr(l)=b} f^{q}_{lt} \notag \\
    & \hspace{95pt} - D^{q}_b - \Delta D^{q+}_{bt} + \Delta D^{q-}_{bt} = 0.
\end{align}
For load buses $b\in\mathcal{N} \setminus \mathcal{N}^{\text{subs}}$:
\begin{align}
    & \sum_{l\in\mathcal{L}|to(l)=b} \left(f^{p+}_{lt} - f^{p-}_{lt}\right) - \sum_{l\in\mathcal{L}|fr(l)=b} \left(f^{p+}_{lt} - f^{p-}_{lt}\right) \notag \\ 
    & \hspace{95pt} -D^{p}_{bt} - \Delta D^{p+}_{bt} + \Delta D^{p-}_{bt} = 0, \\
    & \sum_{l\in\mathcal{L}|to(l)=b} f^{q}_{lt} - \sum_{l\in\mathcal{L}|fr(l)=b} f^{q}_{lt} - D^{q}_b - \Delta D^{q+}_{bt} \notag \\ 
    & \hspace{163pt} + \Delta D^{q-}_{bt} = 0.
\end{align}
For each buses $b\in\mathcal{N}$:
\begin{align}
    & \sum_{l\in\mathcal{L}|to(l)=b} w_{lt}^{\mathrm{op}} + \sum_{l\in\mathcal{L}|fr(l)=b} w_{lt}^{\mathrm{op}} \le M(1-\iota_{bt})\\
    & \sum_{l\in\mathcal{L}|to(l)=b} w_{lt}^{\mathrm{op}} + \sum_{l\in\mathcal{L}|fr(l)=b} w_{lt}^{\mathrm{op}} \ge 1-M\iota_{bt}.
\end{align}
For each switchable line $l\in\mathcal{L}^{\text{sw}}$:
\begin{align}
    & -v_{fr(l),t} + v_{to(l),t} + 2 \left( R_l \cdot \left(f^{p+}_{lt} - f^{p-}_{lt}\right) + X_l \cdot f^{q}_{lt} \right) \notag\\ 
    & \hspace{125pt} - (1 - z_{lt}^{\mathrm{sw}}) \cdot M \leq 0, \\ 
    & v_{fr(l),t} - v_{to(l),t} - 2 \left( R_l \cdot \left(f^{p+}_{lt} - f^{p-}_{lt}\right) + X_l \cdot f^{q}_{lt} \right) \notag \\ 
    & \hspace{125pt} -(1 - z_{lt}^{\mathrm{sw}}) \cdot M \leq 0.
\end{align}
For non-switchable lines $l\in\mathcal{L}\setminus \mathcal{L}^{\text{sw}}$
\begin{align}
    & -v_{fr(l),t} + v_{to(l),t} + 2 \left( R_l \cdot \left(f^{p+}_{lt} - f^{p-}_{lt}\right) + X_l \cdot f^{q}_{lt} \right) \notag\\  
    & \hspace{120pt} - (1 - a^{avail}_{lt}) \cdot M \leq 0, \label{eq:vol_1}\\ 
    & v_{fr(l),t} - v_{to(l),t} - 2 \left( R_l \cdot \left(f^{p+}_{lt} - f^{p-}_{lt}\right) + X_l \cdot f^{q}_{lt} \right) \notag \\ 
    & \hspace{120pt} - (1 - a^{avail}_{lt}) \cdot M \leq 0. \label{eq:vol_2}
\end{align}
For each bus $b \in \mathcal{N}^\text{sub}$:
\begin{align}
    v_{bt} = V_{ref}^2.
\end{align}
For each bus $b \in \mathcal{N}$:
\begin{align}
    & \underline{V_b}^2\le v_{bt} \le \overline{V_b}^2\\
    & \Delta D^{p-}_{bt} \leq D^p_b, \\
    & \Delta D^{q-}_{bt} \leq D^q_b.
\end{align}
For each line $l  \in \mathcal{L}$:
\begin{align}
    & 0\le f^{p+}_{lt} \leq F_{max} \cdot d_{lt}, \\
    & 0\le f^{p-}_{lt} \leq F_{max} \cdot (1 - d_{lt}).
\end{align}
For each line $l \in \mathcal{L}^\text{sw}$:
\begin{align}
    & 0\le f^{p+}_{lt} \leq F_{max} \cdot z_{lt}^{\mathrm{sw}} \\
    & 0\le f^{p-}_{lt} \leq F_{max} \cdot z_{lt}^{\mathrm{sw}} \\
    & -F_{max} \cdot z_{lt}^{\mathrm{sw}} \leq f^{q}_{lt} \leq F_{max} \cdot z_{lt}^{\mathrm{sw}}.
\end{align}
For each line $l \in \mathcal{L}/\mathcal{L}^\text{sw}$:
\begin{align}
    & f^{p+}_{lt} \leq F_{max} \cdot a^{avail}_{lt} \\
    & f^{p-}_{lt} \leq F_{max} \cdot a^{avail}_{lt}.
\end{align}
For $e \in [1,4]$ and $l \in \mathcal{L}$:
\begin{align}
    & f^{q}_{lt} - \cot[{(1/2-e)\pi/4}] \cdot (f^{p+}_{lt} - f^{p-}_{lt} \notag  \\ 
    & \hspace{35pt} -\cos[{e\pi/4}] \cdot F_{max}) -\sin[{e\pi/4}] \cdot F_{max} \leq 0 \\
    & -f^{q}_{lt} - \cot[{(1/2-e)\pi/4}] \cdot (f^{p+}_{lt} - f^{p-}_{lt} \notag  \\ 
    & \hspace{35pt} -\cos[{e\pi/4}] \cdot F_{max}) -\sin[{e\pi/4}] \cdot F_{max} \leq 0.
\end{align}
For each line $l \in \mathcal{L}^\text{sw}$:
\begin{align}
    & y_{lt}^{\mathrm{sw}} \geq z_l^{\mathrm{sw},t-1} - z_{lt}^{\mathrm{sw}} \\
    & y_{lt}^{\mathrm{sw}} \geq z_{lt}^{\mathrm{sw}} - z_l^{\mathrm{sw},t-1}\\
    & z_{lt}^{\mathrm{sw}} \leq a^{avail}_{lt}\\
    & w_{lt}^{\mathrm{op}} = z_{lt}^{\mathrm{sw}}.
\end{align}
To enforce radiality of the network reconfiguration, we adopt the radiality constraints in \cite{babaei2020distributionally} and construct a spanning forest over the energized buses at each time period. Let $s$ denote an artificial super-root node, and connect $s$ to each substation $b\in\mathcal{N}^{\mathrm{subs}}$. Let $\mathcal{A}'$ denote the directed arc set of the augmented graph. For each directed arc $(i,j)\in\mathcal{A}'$, we introduce a nonnegative fictitious flow variable $f_{ijt}^{\mathrm{rad}}$, which is used only to enforce radiality.

For each time period $t$, the super-root injects one unit of fictitious flow for each energized bus:
\begin{align}
\sum_{b\in\mathcal{N}^{\mathrm{subs}}} f_{sbt}^{\mathrm{rad}}
=
\sum_{b\in\mathcal{N}} (1-\iota_{bt}).
\end{align}

For each bus $b\in\mathcal{N}$, flow conservation is imposed as
\begin{align}
\sum_{i:(i,b)\in\mathcal{A}'} f_{ibt}^{\mathrm{rad},t}
-
\sum_{j:(b,j)\in\mathcal{A}'} f_{bjt}^{\mathrm{rad},t}
=
1-\iota_{bt}.
\end{align}

For each real line $l=\{i,j\}\in\mathcal{L}$, fictitious flow is allowed only if the line is in operation:
\begin{align}
f_{ij}^{\mathrm{rad},t} \le M^{\mathrm{rad}} w_{lt}^{\mathrm{op}},\\
f_{ji}^{\mathrm{rad},t} \le M^{\mathrm{rad}} w_{lt}^{\mathrm{op}},
\end{align}
where $M^{\mathrm{rad}}$ is a sufficiently large constant, e.g., $M^{\mathrm{rad}}=|\mathcal{N}|$.

Finally, the number of energized real lines must equal the number of energized buses minus the number of substations:
\begin{align}
\sum_{l\in\mathcal{L}} w_{lt}^{\mathrm{op}}
=
\sum_{b \in \mathcal{N} \setminus \mathcal{N}^{\text{subs}}} (1-\iota_{bt}) .
\end{align}
These constraints ensure that the energized portion of the network forms a radial forest rooted at the substations, while isolated buses are admissible.

\vspace{-0.4cm}
\subsection{Approximate Dynamic Programming}
To solve the multi-period DRMDP efficiently, we employ an ADP algorithm with linear value-function approximation. The key idea is to represent the value function 
\begin{equation}
    V_t(\tilde{\boldsymbol{s}})
\approx
\boldsymbol{\theta}_t^\top 
\boldsymbol{\phi}(\tilde{\boldsymbol{s}}),
\end{equation}
where $\tilde{\boldsymbol{s}}$ denotes the post-decision state, i.e., the system state immediately after applying the switching action but before uncertainty is realized. $\boldsymbol{\phi}(\tilde{\boldsymbol{s}})$ is a feature vector, and $\boldsymbol{\theta}_t$ is a learned parameter vector. 

In our implementation, $\boldsymbol{\phi}(\tilde{\boldsymbol{s}})$ encodes the operational status of each line in the wildfire zone as a binary tuple. For each line $l\in\mathcal{L}^{fr}$, the corresponding feature entry equals
\begin{equation}
    \phi_l(\tilde{\boldsymbol{s}}) = a^{avail}_{lt}\cdot y^{\text{sw}}_{lt} \in \{0,1\},
\end{equation}
where $\phi_l(\tilde{\boldsymbol{s}})=1$ only when the line is both available and switched on; otherwise $\phi_l(\tilde{\boldsymbol{s}})=0$, capturing cases where the line is either unavailable due to wildfire damage or intentionally de-energized. Using post-decision states separates the deterministic impact of switching actions from subsequent stochastic evolution, which simplifies the Bellman recursion and enables a tractable approximation of multi-stage decision making \cite{wang2019markov}.

At each iteration, the algorithm simulates multiple trajectories of network evolution. For each state $\boldsymbol{s}_t$ encountered, the robust one-step value is computed as
\begin{equation}
    Q_{\boldsymbol{a}\boldsymbol{s},t} = r_{\boldsymbol{a}\boldsymbol{s},t} + \lambda \min_{i=1,\dots,N} \sum_{\boldsymbol{s}'} p_{\boldsymbol{a}\boldsymbol{s},it}(\boldsymbol{s}')\, \boldsymbol{\theta}_{t+1}^{\top} \boldsymbol{\phi}(\boldsymbol{s}')
\end{equation}
and the greedy action $\boldsymbol{a}_t^{*} = \arg\max_{\boldsymbol{a}\in\mathcal{A}(\boldsymbol{s})} Q_{\boldsymbol{a}\boldsymbol{s},t}$ is selected.   

The resulting post-decision state $\boldsymbol{s}_t^{a}$ and its value $Q^{t*}_{\boldsymbol{a} \boldsymbol{s} }$ are stored.  
The next state $\boldsymbol{s}_{t+1}$ is then sampled using simulated line-failure outcomes with $\varepsilon$-greedy exploration to ensure broad state-space coverage.  
After all trajectories are generated, the value-function parameters are updated via ridge-regularized least squares:
\begin{equation}
\boldsymbol{\theta}_t
=\arg\min_{\boldsymbol{\theta}}
\sum_{(\boldsymbol{\phi}, \nu) \in \mathcal{J}_t}
\left(\nu - \boldsymbol{\theta}^\top \boldsymbol{\phi}
\right)^2+\eta \|\boldsymbol{\theta}\|_2^2.
\end{equation}
Here, $\mathcal{J}_t$ denotes the set of sample pairs $(\boldsymbol{\phi},\nu)$ collected at stage $t$, and $\eta>0$ is a regularization parameter that controls overfitting and stabilizes the parameter estimates. The outer loop repeats until the parameter sequence $\{\boldsymbol{\theta}_t\}_{t=1}^T$ converges.
\begin{algorithm}
\caption{Linear ADP for the DRMDP}
\label{alg:LFA_ADP}
\footnotesize
\begin{enumerate}
    \item Initialize value-function parameters $\{\boldsymbol{\theta}_t^{(0)}\}_{t=1}^{T-1}$.
    \item \textbf{For} outer iterations $n = 1, \dots, N_{\text{outer}}$:
    \begin{enumerate}
        \item Set sample buffers $\mathcal{J}_t \leftarrow \varnothing$ for all $t$.
        \item \textbf{Trajectory simulation:} for $m = 1, \dots, M$:
        \begin{enumerate}
            \item Sample initial state $\boldsymbol{s}_1^{(m)} \sim \mu_0$.
            \item \textbf{For} $t = 1, \dots, T-1$:
            \begin{enumerate}
                \item Compute
                    \[
                    \begin{aligned}
                    Q_{\boldsymbol{a}\boldsymbol{s},t}&= r_{\boldsymbol{a}\boldsymbol{s},t}\\
                    &+ \lambda \min_{i=1,\dots,N}
                    \sum_{\boldsymbol{s}'\in\mathcal{S}}
                    p_{\boldsymbol{a}\boldsymbol{s},it}(\boldsymbol{s}')\,
                    \boldsymbol{\theta}_{t+1}^{(n-1)\top}\,\boldsymbol{\phi}(\boldsymbol{s}')
                    \end{aligned}
                    \]
                \item Select greedy action
                $\boldsymbol{a}_t^{*}=\arg\max_{\boldsymbol{a}\in\mathcal{A}(\boldsymbol{s})}
\; Q_{\boldsymbol{a}\boldsymbol{s},t}$
                and set $\nu_t = Q^{*}_{\boldsymbol{a} \boldsymbol{s},t}$.
                \item Form post-decision state $\boldsymbol{s}_t^a$ and add
                $(\boldsymbol{\phi}(\tilde{\boldsymbol{s}_t}), \nu_t)$ to buffer $\mathcal{J}_t$.
                \item Sample next state $\boldsymbol{s}_{t+1}$ using
                $\varepsilon$-greedy transition sampling.
            \end{enumerate}
        \end{enumerate}
        \item \textbf{Parameter update:} for $t = 1, \dots, T-1$:
        \begin{enumerate}
            \item Build $\boldsymbol{X}_t$ and $\boldsymbol{y}_t$ from $\mathcal{J}_t$.
            \item Update
            \[
              \boldsymbol{\theta}_t^{(n)}=\left(\boldsymbol{X}_t^\top \boldsymbol{X}_t + \eta\,\boldsymbol{I}\right)^{-1}\boldsymbol{X}_t^\top \boldsymbol{y}_t.
            \]
        \end{enumerate}
        \item \textbf{If}
        $\max_t \|\boldsymbol{\theta}_t^{(n)} - \boldsymbol{\theta}_t^{(n-1)}\|_\infty < \texttt{tol}$, \textbf{stop}.
    \end{enumerate}
\end{enumerate}

\textbf{Output:} parameters $\{\boldsymbol{\theta}_t^{(*)}\}$ and greedy policy
\[
\pi^{*}(\boldsymbol{s},t)=\arg\max_{\boldsymbol{a}\in\mathcal{A}}
\Biggl\{r_{\boldsymbol{a}\boldsymbol{s}}+\lambda\,
\min_{i\in\{1,\dots,N\}}
\sum_{\boldsymbol{s}'\in\mathcal{S}}
p_{\boldsymbol{a}\boldsymbol{s},it}(\boldsymbol{s}')
\,
\boldsymbol{\theta}_{t+1}^{*\top}\,
\boldsymbol{\phi}(\boldsymbol{s}')
\Biggr\}.
\]
\end{algorithm}

To ensure broad generalization, each trajectory starts from a randomly sampled feasible state, and the simulation includes exploration via random transitions with high probability. This ensures the value function learns across diverse and rare configurations.

According to \cite{wang2019markov}, post-decision states effectively separate immediate decisions from downstream uncertainty. As a result, each decision problem becomes one-period deterministic, enabling tractable approximation of robust multi-stage control under uncertainty.



\vspace{-0.5cm}
\section{Case Studies}
\vspace{-0.2cm}
To assess the performance of the proposed DRMDP–ADP framework, we conduct two case studies using the 54-bus and 138-bus distribution test systems in \cite{moreira2024distribution}. Both networks include switchable lines as well as wildfire-prone lines, making them well-suited for evaluating decision-making under wildfire-induced uncertainty. We set the energy price to \$0.01/kWh and assign a \$100 cost to each switching action. We benchmark our method against two baselines: (i) a non-decision-dependent uncertainty (non-DDU) model, and (ii) a greedy myopic strategy. All methods are evaluated via an extensive out-of-sample Monte Carlo simulation. 

Our experimental setting follows \cite{moreira2023dataset} except for the line-availability parameters $\gamma_l$ and $\beta_l$ in the transition model (Section~\ref{sec:model}). To reflect spatial heterogeneity in wildfire exposure, we apply a larger $\beta_l$ to lines in the wildfire zone ($l \in \mathcal{L}^{fr}$) and a much smaller $\beta_l$ to lines outside the zone ($l \notin \mathcal{L}^{fr}$). The parameter values used for the 54-bus and 138-bus systems are summarized in Table~\ref{tab:line_availability_params}.
\vspace{-0.2cm}
\begin{table}[htbp]
    \centering
    \footnotesize
    \caption{Line-Availability Parameters Used in the Case Studies}
    \label{tab:line_availability_params}
    \renewcommand{\arraystretch}{1.15}
    \begin{tabular}{lccc}
        \toprule
        \textbf{System} & $\gamma_l$ & $\beta_l \ (l \in \mathcal{L}^{\text{risk}})$ & $\beta_l \ (l \notin \mathcal{L}^{\text{risk}})$ \\
        \midrule
        54-bus  & 0.9989 & 3.0000 & 0.0001 \\
        138-bus & 0.9996 & 1.0005 & 0.0014 \\
        \bottomrule
    \end{tabular}
\end{table}
\vspace{-0.8cm}
\subsection{Experiment Design}
\vspace{-0.2cm}
For each method, we first solve the multi-period distribution system operation problem under its respective modeling assumptions. Then, based on the resulting switching decisions and power flows, we compute failure probabilities for each line using the wildfire-aware transition model in \eqref{eq:tran_prob}. We simulate line failures via 1,000 independent Bernoulli trials across a 20-hour horizon, totaling 20,000 network instances per method.


To capture varying wildfire intensities, we define $N=3$ candidate transition distributions by varying the sensitivity parameter $\beta_l$, following the structure introduced in Section~\ref{sec:trans_prob}. These distributions, corresponding to low-, nominal-, and high-risk wildfire scenarios, collectively form a discrete uncertainty set over $\left\{ p_{\boldsymbol{a}\boldsymbol{s},i}(\cdot) \right\}_{i=1}^{N}$, which is used to construct the ambiguity set in the DRMDP framework.

The different risk levels are modeled by shifting the baseline $\beta_l$ values by $\delta=0.5$ for lines located within wildfire-prone zones, while keeping $\beta$ fixed at a lower value for lines outside these zones. This setup enables a spatially heterogeneous and operationally relevant representation of fire-induced failure risk. As a result, the model can evaluate each method’s performance under diverse and physically realistic stress conditions, enabling robust comparison across the 54-node and 138-node systems.

\paragraph{\bf Decision-Dependent Uncertainty (DDU) Scenario}
In the solving stage, we employ the ADP algorithm to approximate the value function in~\eqref{eq:dual_constraint}, allowing the model to fully capture how switching actions influence future wildfire-induced failure probabilities.  After convergence of the ADP iterations, the learned parameters $\boldsymbol{\theta}_t^{(*)}$ transform the multi-period MDP in Section~\ref{sec:DSO} into a sequence of independent one-period deterministic problems, which can be solved efficiently at runtime.  
During out-of-sample evaluation, successor states are generated using the worst-case transition distribution, reflecting how the policy performs under the most adverse wildfire conditions.

\paragraph{\bf Non-DDU Baseline}
In this version, we disable the decision dependence of the ambiguity set defined in Section \ref{sec:DDU} by setting $\beta_l = 0, \forall l\in\mathcal{L}$, thereby assuming that switching actions have no impact on line failure probabilities in \eqref{eq:tran_prob}.
During out-of-sample simulation, however, the true $\beta_l$ values (identical to those used in the DDU case) are reintroduced.  
This allows us to quantify the performance degradation caused by ignoring endogenous effects in the transition probabilities.

\paragraph{\bf Greedy Baseline}
The greedy strategy serves as a myopic benchmark without look-ahead or value-function approximation.  
We eliminate the recursive term $\alpha$ in~\eqref{eq:recur_obj} and drop the dual constraint~\eqref{eq:dual_constraint}.  
At each time step, a single-period deterministic optimization problem containing all operational constraints in Section~\ref{sec:oper} is solved to minimize only the immediate cost.  
This baseline showcases the consequences of ignoring multi-period coupling and future wildfire risk evolution.
\vspace{-0.3cm}
\subsection{54-Node System}
\vspace{-0.2cm}
Table~\ref{tab:54_system} compares the performance of the DDU, Non-DDU, and Greedy strategies under both average and worst-case wildfire scenarios across 1,000 simulations. 
Under average-case conditions, the DDU method achieves the lowest total cost, significantly outperforming both the Non-DDU and Greedy baselines. 
In the worst 5\% scenarios, 
the DDU policy remains the most resilient, achieving the lowest total cost along with reduced load loss and fewer line failures relative to the Non-DDU and Greedy baselines. 

\begin{table*}[htbp]
\centering
\caption{Operational Performance Comparison Under Average and Worst-Case Wildfire Scenarios for the 54-Node System}
\label{tab:54_system}
\renewcommand{\arraystretch}{1.2}
\begin{tabular}{lcccccc}
\toprule
\multirow{2}{*}{\textbf{Metric}} & \multicolumn{3}{c}{\textbf{Average accross scenarios}} & \multicolumn{3}{c}{\textbf{Worst 5\% scenarios}} \\
\cmidrule(r){2-4} \cmidrule{5-7}
& \textbf{DDU} & \textbf{Non-DDU} & \textbf{Greedy} & \textbf{DDU} & \textbf{Non-DDU} & \textbf{Greedy} \\
\midrule
Total Cost (\$)            & 356.93 & 500.17 & 509.56 & 1821.76 & 2000.34 & 2040.53 \\
Power Purchase Cost (\$)   & 53.64  & 53.46  & 53.45  & 51.83   & 51.61   & 51.56   \\
Switching Cost (\$)        & 16.98  & 16.86  & 17.43  & 25.30   & 27.80   & 29.60   \\
Load Loss Cost (\$)        & 286.31 & 429.85 & 438.67 & 1744.62 & 1920.93 & 1959.37 \\
No. of Failed Lines        & 1.67   & 2.70   & 2.77   & 4.2      & 5.4      & 5.4      \\
Load Shedding (MW, 20-hr)        & 28.63 & 42.99 & 43.87 & 174.46 & 192.09 & 195.94 \\
Highest Hourly Load Shedding (\% of demand)  & 0.76 & 1.18 & 1.21 & 4.83 & 5.56 & 5.58 \\
\bottomrule
\end{tabular}
\end{table*}

\begin{figure}[!ht]
\centering
\includegraphics[width=\linewidth]{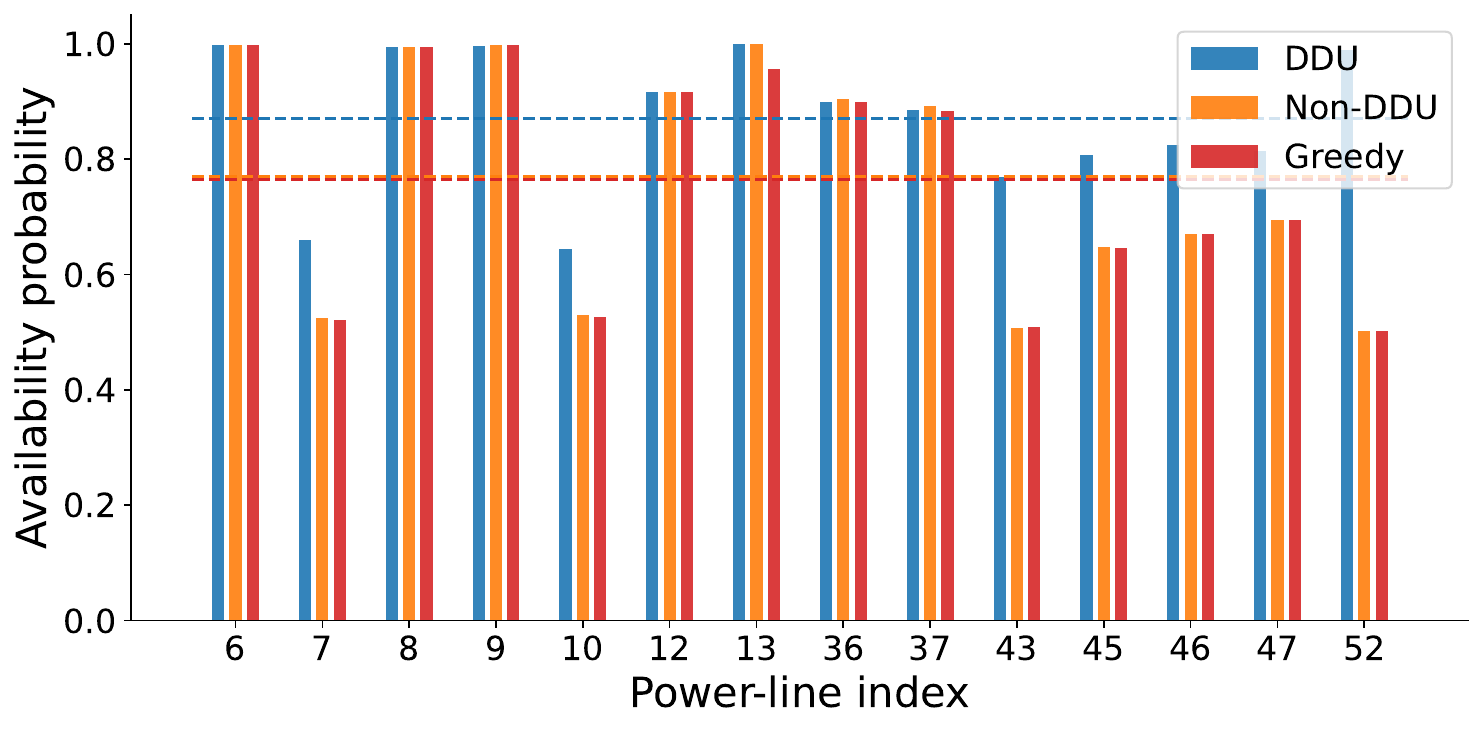}
\caption{Line availability probability for the 54-bus network across 1000 scenarios among all methods.}
\label{fig:54-avl}
\end{figure}

Fig.~\ref{fig:54-avl} presents the average availability probabilities for a subset of wildfire-prone lines in the 54-bus network. Each bar represents the average availability of a specific line over 1,000 simulated wildfire scenarios. As shown, the DDU method consistently yields higher availability probabilities across nearly all monitored lines. This improvement stems from its proactive switching behavior, which strategically reduces power flows through vulnerable lines, thereby decreasing thermal and ignition-related failure risks. In contrast, the Greedy method results in the lowest line availability, reflecting its lack of long-term risk mitigation. By minimizing only immediate operational cost, the Greedy policy permits higher sustained loading on critical lines, which elevates their probability of failure under adverse conditions.
\vspace{-0.1cm}

\begin{figure}[!ht]
\centering
\includegraphics[width=\linewidth]{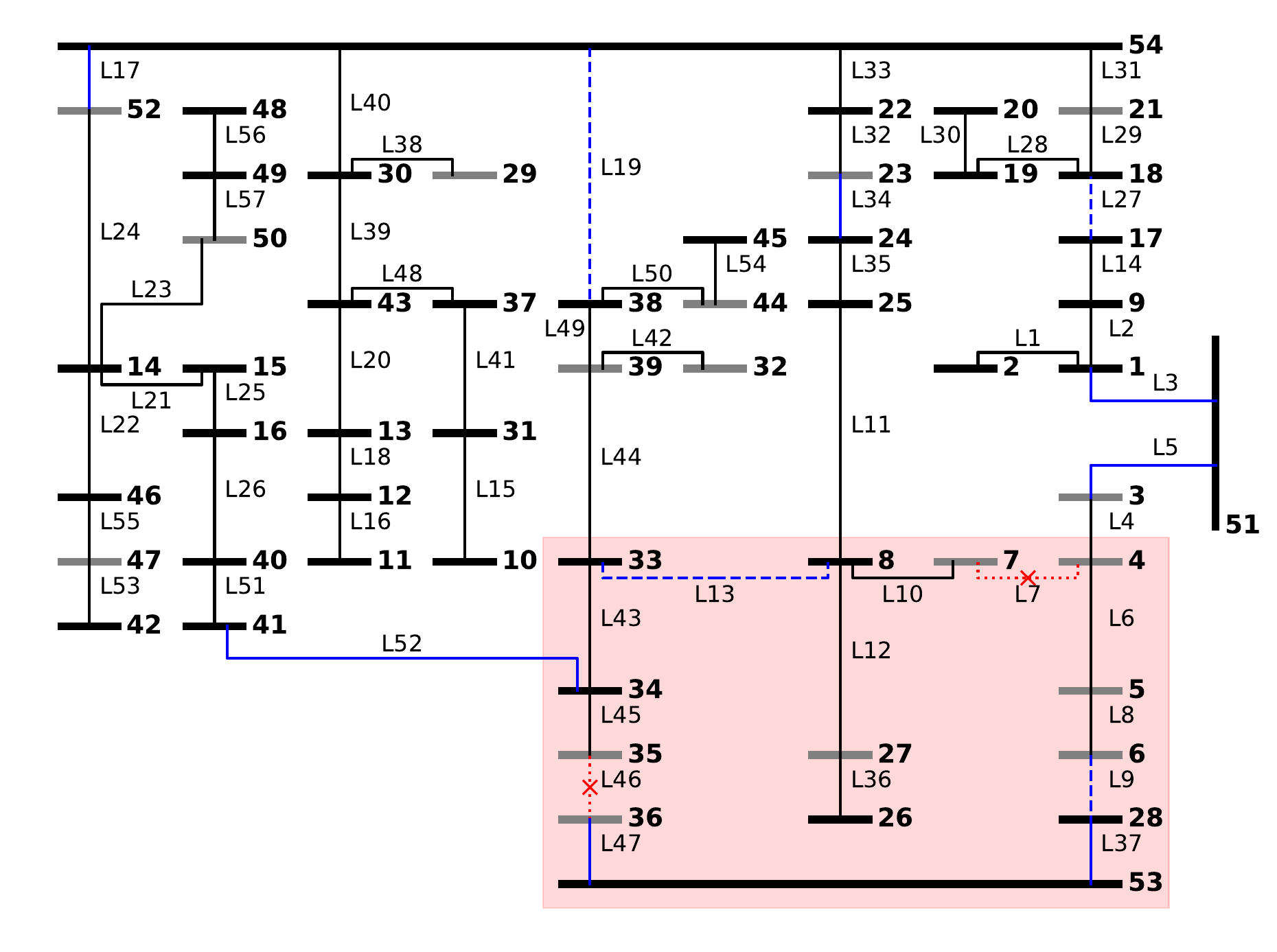}
\caption{Final network topology under the DDU method at hour 20 in representative scenario ID 0 of the 54-bus network.}
\label{fig:DDU-54}
\end{figure}  

\begin{figure}[!ht]
\centering
\includegraphics[width=\linewidth]{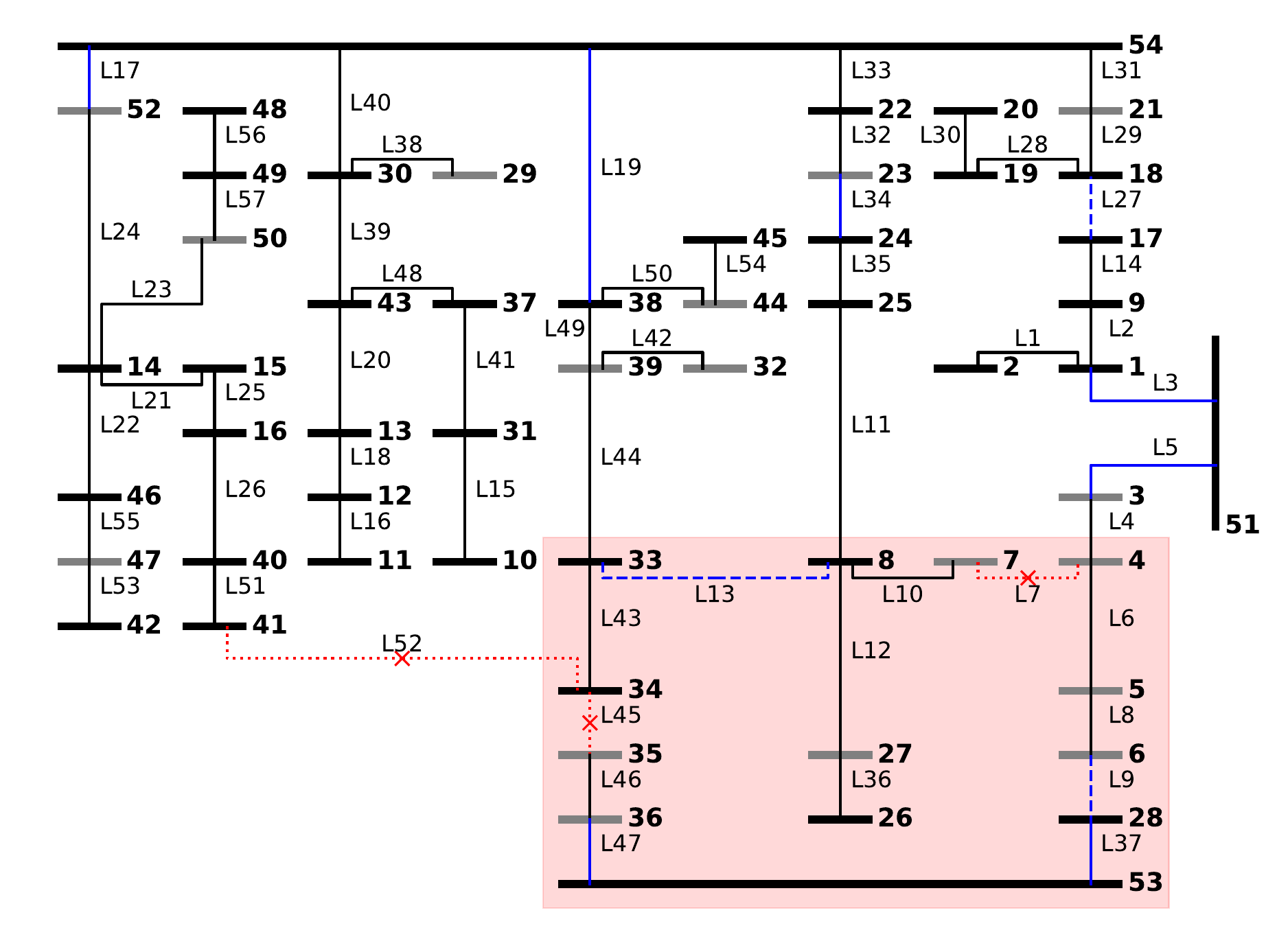}
\caption{Final network topology under the Non-DDU and Greedy methods at hour 20 in scenario ID 0 of the 54-bus network.}
\label{fig:noDDU-54}
\end{figure}  
In a representative wildfire scenario (Scenario 0), the switching policies and resulting topologies for the three methods highlight clear differences in detail. As shown in Table~\ref{tab:switching_summary}, both the Greedy and Non-DDU approaches follow the same sequence of actions: switching on L19 at Hour 2, L34 at Hour 3, and L17 at Hour 4. These decisions are made reactively, only after failures have occurred (e.g., L45, L7, and L43), resulting in three failed lines and a substantial load loss cost of \$680. The corresponding final topology in Fig.~\ref{fig:noDDU-54} reveals that power restoration in the fire zone relies heavily on late-stage interventions that are unable to prevent cascading failures.

In contrast, the DDU method demonstrates a more anticipatory strategy. It proactively switches on L17 as early as Hour 1, prior to any line failure, and then activates L19 and L34 in subsequent hours. This early reconfiguration helps redistribute power flows and reduce stress on vulnerable lines. As illustrated in Fig.~\ref{fig:DDU-54}, this policy successfully prevents the failure of L43, leading to only two failed lines and eliminating all load loss penalties. The outcome underscores that DDU is able to foresee the consequences of switching on future failure probabilities and adapts accordingly, enabling more robust and cost-effective operation under wildfire conditions.


\begin{table}[htbp]
\centering
\caption{Switching Actions and Consequences for a Representative Wildfire Scenario for the 54-Node System}
\label{tab:switching_summary}
\renewcommand{\arraystretch}{1.3}
\begin{tabular}{l p{3cm} c}
\toprule
\textbf{Method} & \textbf{Switching Actions} &  \textbf{Total Cost} \\
\midrule
Greedy & 
Hour 2: L34 on [L7 fail] \newline
Hour 3: L17 on [L52 fail] \newline
Hour 4: L19 on [L52 fail]
 & 74.00 \\
\midrule
Non-DDU & 
Hour 2: L34 on [L7 fail] \newline
Hour 3: L17 on [L52 fail] \newline
Hour 4: L19 on [L45 fail]
 & 75.43 \\
\midrule
DDU & 
Hour 1: L17 on \newline
Hour 2: L34 on [L7 fail] \newline
Hour 3: [L46 fail]
 & 64.00 \\
\bottomrule
\end{tabular}
\end{table}

\begin{figure}[!ht]
\centering
\includegraphics[width=\linewidth]{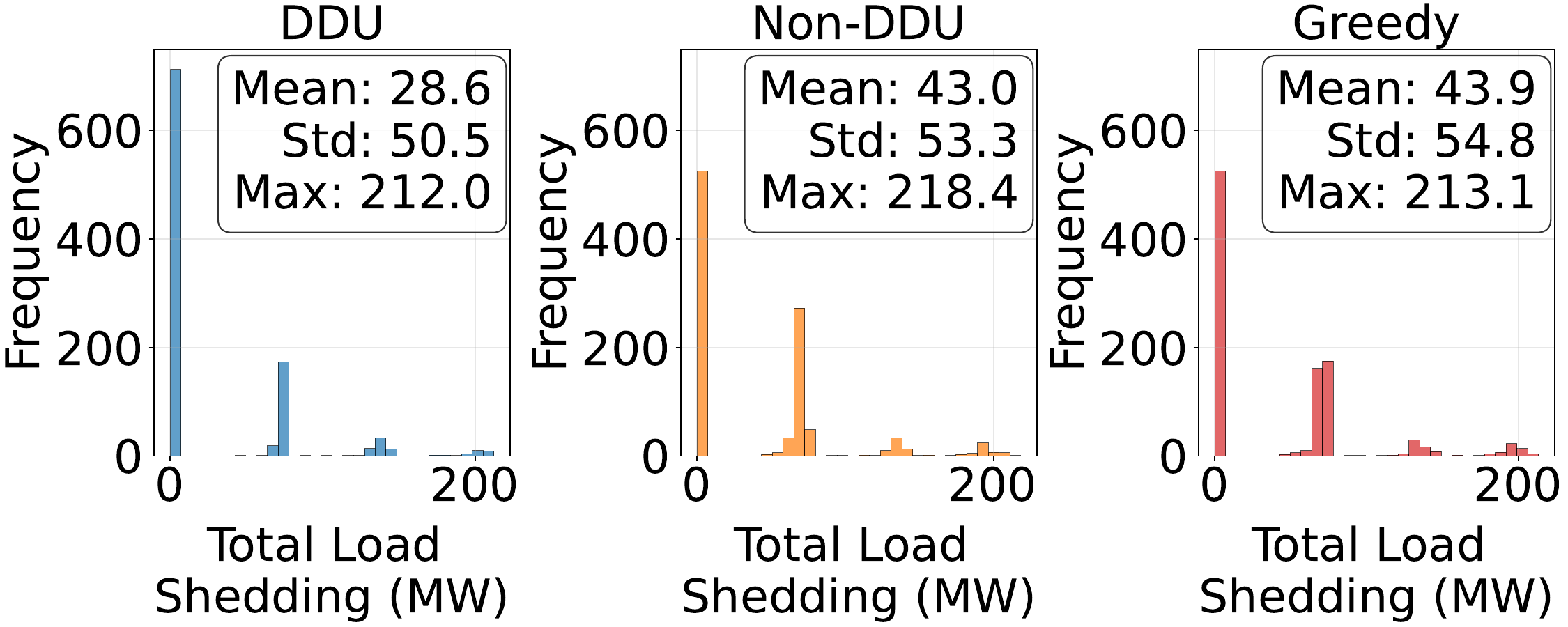}
\caption{Distribution of total daily load shedding over 20-hour simulation across all methods for the 54-Node System.}
\label{fig:54-his}
\end{figure}  
Fig.~\ref{fig:54-his} presents the distribution of hourly load shedding amounts over 1,000 wildfire scenarios for each method in the 54-node system. The DDU strategy yields a sharply concentrated distribution around zero, with 71.2\% of hourly instances experiencing no load shedding, substantially higher than the 52.5\% observed under both the Non-DDU and Greedy methods. In contrast, both the Non-DDU and Greedy strategies show heavier tails, indicating a higher frequency of severe outages and costlier disruptions. Notably, the DDU method achieves the lowest mean and standard deviation for the load shedding, along with a slightly lower maximum, highlighting its ability to reduce both the average and variability of wildfire-related service interruptions. These results reinforce that anticipatory switching under decision-dependent uncertainty not only improves average performance but also provides more consistent resilience across a wide range of stochastic conditions.
\vspace{-0.4cm}
\subsection{138-Node System}

\begin{table*}[t]
\centering
\caption{Operational Performance Comparison Under Average and Worst-Case Wildfire Scenarios for the 138-Node System}
\label{tab:138_system}
\renewcommand{\arraystretch}{1.2}
\begin{tabular}{lcccccc}
\toprule
\multirow{2}{*}{\textbf{Metric}} & \multicolumn{3}{c}{\textbf{Average across scenarios}} & \multicolumn{3}{c}{\textbf{Worst 5\% scenarios}} \\
\cmidrule(r){2-4} \cmidrule{5-7}
& \textbf{DDU} & \textbf{Non-DDU} & \textbf{Greedy} & \textbf{DDU} & \textbf{Non-DDU} & \textbf{Greedy} \\
\midrule
Total Cost (\$)            & 7561.33 & 9464.53 & 9464.36 & 18307.93 & 18919.76 & 18919.66 \\
Power Purchase Cost (\$)   & 560.27  & 557.88  & 557.88  & 546.83  & 546.06   & 546.06   \\
Switching Cost (\$)        & 16.90  & 14.57  & 14.40  & 23.50   & 19.30   & 19.20   \\
Load Loss Cost (\$)        & 6984.16 & 8892.08 & 8892.08 & 17737.60 & 18354.40 & 18354.40 \\
No. of Failed Lines        & 2.42   & 3.77   & 3.77   & 6.6      & 5.3      & 5.3      \\
Load Shedding (MW, 20-hr)      & 698.42 & 889.21 & 889.21 & 1773.76 & 1835.44 & 1835.44 \\
Highest Hourly Load Shedding (\% of demand)  & 19.93 & 25.83 & 25.83 & 52.70 & 53.00 & 53.00 \\
\bottomrule
\end{tabular}
\end{table*}
\vspace{-0.1cm}
Table~\ref{tab:138_system} illustrates the operational performance of the three methods on the 138-node system under both average and worst-case wildfire scenarios. The DDU method consistently outperforms both Non-DDU and Greedy baselines, achieving the lowest total cost across both settings. Despite incurring slightly higher switching costs, DDU demonstrates its robustness by cutting load loss. These results highlight the advantage of proactive and risk-aware decision-making.

\begin{figure}[!ht]
    \centering
    \includegraphics[width=\linewidth]{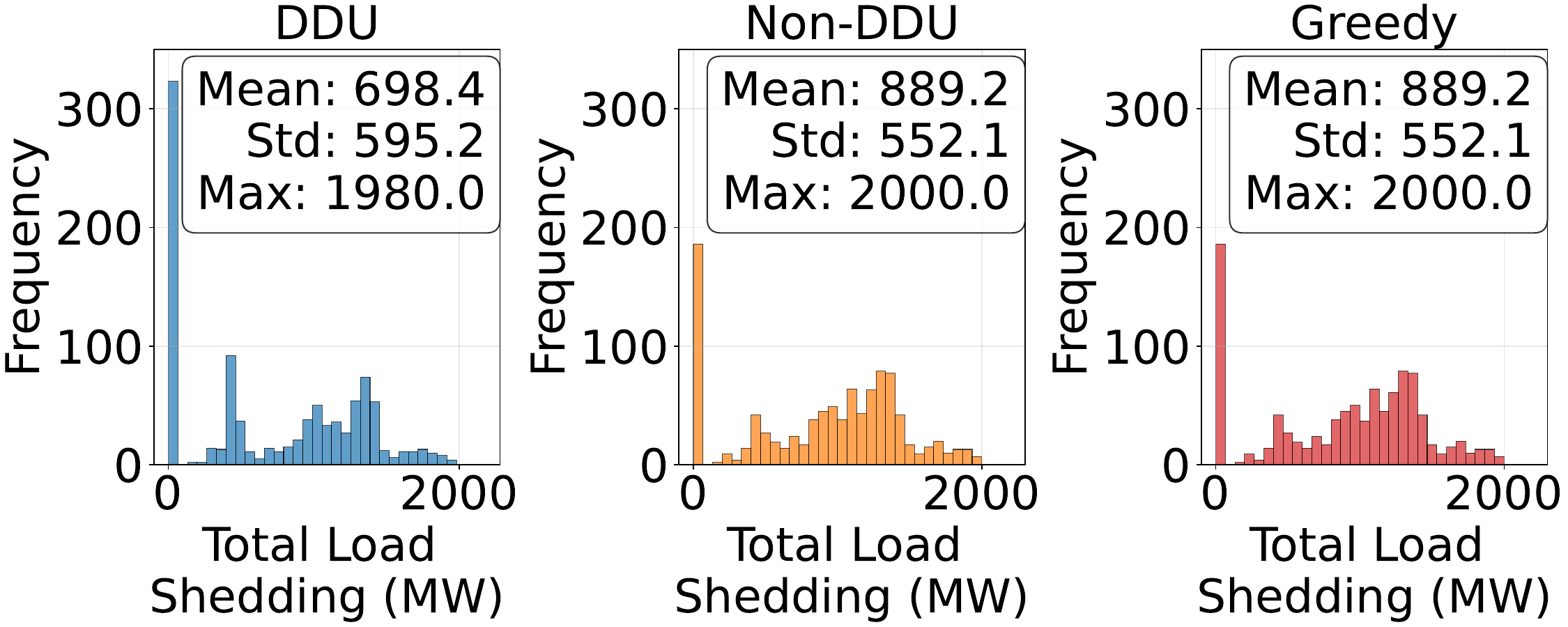}
    \caption{Distribution of total daily load shedding over 20-hour simulation across all methods for the 138-Node System.}
    \label{fig:138-his}
\end{figure}

Fig. \ref{fig:138-his} illustrates the distribution of hourly load shedding across 1,000 wildfire scenarios for the 138-node system. Compared to Fig. \ref{fig:54-his}, the DDU method exhibits a wider spread of outcomes due to the increased system complexity. Nevertheless, it still maintains a notable concentration near zero, with 32.2\% of hourly instances exhibiting no load loss, significantly higher than the 18.6\% observed for both the Non-DDU and Greedy methods. 

Despite requiring moderately longer computation time, the DDU method consistently delivers superior operational performance. As reported in Table~\ref{tab:computation}, the runtime of DDU includes an offline training phase in which the ADP algorithm iteratively simulates trajectories and fits the time-indexed value-function approximations until convergence. This training step explains the slightly higher runtime observed for the 54-bus system. Importantly, the ADP training can be performed offline and amortized over repeated operations once the value functions $\{\boldsymbol{\theta}_t^{(*)}\}$ are learned.

To interpret Table~\ref{tab:computation} in an operational setting where operation decisions are made hourly, we note that the reported runtimes correspond to solving the full 20-hour horizon (including ADP training for DDU). A conservative estimate of the average wall-clock time per hourly decision can be obtained by dividing the total runtime by 20, which yields approximately 43 s per hourly run for the 54-bus system and 109 s for the 138-bus system under DDU. In practice, once $\{\boldsymbol{\theta}_t^{(*)}\}$ are learned, the operator would only incur the per-hour solve time of the single-period model, making the proposed approach suitable for near-real-time use while delivering substantial reliability and cost benefits.
Hence, the marginal computational burden during real-time operation is comparable to the baselines, while retaining the resilience and cost advantages of decision-dependent uncertainty modeling.

\begin{table}[htbp]
    \centering
    \caption{Online and Offline Computation Time for 20-Hour Simulations on 54-Bus and 138-Bus Systems}
    \label{tab:computation}
    \resizebox{0.5\textwidth}{!}{
    \begin{tabular}{lcccccc}
        \toprule
        \multirow{2}{*}{\textbf{System}} &
        \multicolumn{3}{c}{\textbf{Online inference (s)}} &
        \multicolumn{3}{c}{\textbf{Offline training (s)}} \\
        \cmidrule(lr){2-4}\cmidrule(lr){5-7}
        & \textbf{DDU} & \textbf{Non-DDU} & \textbf{Greedy}
        & \textbf{DDU} & \textbf{Non-DDU} & \textbf{Greedy} \\
        \midrule
        54-Bus  & 864.94  & 830.65  & 711.85  & 100.31 & 114.69 & -- \\
        138-Bus & 2173.26 & 2409.92 & 1916.44 & 155.30 & 149.45 & -- \\
        \bottomrule
    \end{tabular}}
\end{table}
\vspace{-0.6cm}
\vspace{-0.6cm}
\section{Conclusion}
\vspace{-0.2cm}
This paper presents a novel distributionally robust Markov decision process (DRMDP) framework for wildfire-aware distribution system operation. By explicitly modeling the decision-dependent transition probabilities of power line availability under wildfire risk, the proposed approach captures both exogenous hazard exposure and endogenous uncertainty induced by switching decisions. To address the inherent complexity of multi-stage stochastic optimization under uncertainty, we integrate a linear function approximation-based approximate dynamic programming (ADP) algorithm. This allows for real-time policy generation that accounts for the evolving interplay between grid configuration and future failure risk.

Extensive case studies on the 54-bus and 138-bus systems validate the effectiveness of the proposed framework. Relative to baseline strategies that ignore or simplify the transition dynamics, the DDU-based policy achieves substantially lower total operational cost and load loss under both average conditions and high-impact (worst 5\%) wildfire scenarios. These gains are driven by proactive, risk-aware switching that mitigates downstream outages, while maintaining computational runtimes comparable to the baselines. Overall, the results underscore the importance of explicitly modeling decision-dependent uncertainty for resilient and cost-effective grid operation under wildfire risk.
\vspace{-0.4cm}
\section*{Acknowledgment}
This work has been funded by the U.S. Department of Energy, Office of Electricity, under the contract DE-AC02-05CH11231 and National Science Foundation \#2338559. 
\vspace{-0.4cm}
\bibliography{scholar}
\bibliographystyle{IEEEtran}


 




\vfill

\end{document}